\documentclass[twocolumn,superscriptaddress]{revtex4-2}
\usepackage[T1]{fontenc}
\usepackage{pifont}

\usepackage[ruled,vlined]{algorithm2e}

\usepackage[colorlinks=true,urlcolor=black,citecolor=blue,linkcolor=blue]{hyperref}

\usepackage{mathtools}
\usepackage{amsthm}
\usepackage{amssymb}
\usepackage{graphicx}
\usepackage{booktabs}
\usepackage[dvipsnames]{xcolor}
\usepackage{enumitem}

\graphicspath{figures}

\begin{document}
\title{Forecasting Quantum Observables: \\A Compressed Sensing Approach with Performance Guarantees}
\author{V\'ictor Valls}
\affiliation{IBM Quantum, IBM Research}

\author{Albert Akhriev}

\affiliation{IBM Quantum, IBM Research}

\author{Olatz Sanz Larrarte}
\affiliation{Department of Basic Sciences, Tecnun - University of Navarra, 20018 San Sebastian, Spain}

\author{Javier Oliva del Moral}
\affiliation{Department of Basic Sciences, Tecnun - University of Navarra, 20018 San Sebastian, Spain}
\affiliation{Donostia International Physics Center, 20018 San Sebastián, Spain}

\author{{\v{S}}t{\v{e}}p{\'a}n {\v{S}}m{\'\i}d}
\affiliation{Department of Computing, Imperial College London, UK}
\affiliation{IBM Quantum, IBM Research}

\author{Josu Etxezarreta Martinez}
\affiliation{Department of Basic Sciences, Tecnun - University of Navarra, 20018 San Sebastian, Spain}

\author{Sergiy Zhuk}
\affiliation{IBM Quantum, IBM Research}

\author{Dmytro Mishagli}
\affiliation{IBM Quantum, IBM Research}

\begin{abstract}
Data-driven extrapolation methods aim to extend the dynamics of quantum observables from measurements, but they often lack guarantees on prediction accuracy. We introduce a framework based on atomic norm minimization that can certify whether the spectral model learned by a forecasting algorithm---i.e., Bohr frequencies and amplitudes---is consistent with unitary quantum time evolution. Certification holds when the dynamics are governed by a small number of well-separated Bohr frequencies. We validate the approach on multiple forecasting algorithms applied to spin-chain Hamiltonians with 8--20 sites. Comparing with exact diagonalization, certified models yield an average forecasting error below 0.1 (observable range $[-1, 1]$) in 97\% of cases and below 0.05 in 91--99\% of cases. Even in the presence of noise, certified models remain robust at the 0.1 error threshold.
\end{abstract}

\maketitle

\section{Introduction}

Quantum spin Hamiltonians are natural testbeds for near-term quantum devices because their local interaction structure maps directly onto existing qubit connectivity~\cite{Fauseweh2024}. Recent experiments have already probed dynamics with more than one hundred qubits and circuit depths of several hundred two-qubit gates~\cite{kim2023_evidence_utility_quantum_computing_fault,switzer2025_realization_twodimensional_discrete_time_crystals}. However, the time window for reliable quantum time evolution remains limited: gate errors accumulate rapidly \cite{temme2017error}, and error mitigation is effective only for short times, as its cost grows exponentially \cite{kim2023_evidence_utility_quantum_computing_fault}.

To overcome this limitation, a growing body of work seeks to infer long-time quantum dynamics from early, low-noise measurements~\cite{erpenbeck2025compact,manos2025_extrapolation_quantum_measurement_data,kaneko2025forecasting,shen2024efficient}. These \emph{data-driven} approaches exploit the fact that expectation values can be expressed as the sum of harmonics, enabling extrapolation that is independent of the system size and qubit connectivity. Methods such as ESPRIT and DMD leverage this structure and have shown that they can extrapolate observables well beyond the measured time window~\cite{erpenbeck2025compact,kaneko2025forecasting}. 

Despite their empirical success, existing forecasting methods cannot certify that the extrapolated observables are consistent with the underlying dynamics, i.e., whether the spectral model captures the Bohr frequencies and amplitudes that govern the unitary time evolution. In this work, rather than proposing a forecasting algorithm and assessing its accuracy empirically, we ask a more fundamental question: whether a proposed spectral model captures the underlying unitary quantum time evolution dynamics. Addressing this gap requires moving beyond forecasting alone and introducing a systematic framework for validating candidate models.

In this paper, we introduce such a framework using tools from Atomic Norm Minimization (ANM)~\cite{bhaskar2013atomic,chandrasekaran2012convex}, a compressed sensing approach for sparse spectral recovery with rigorous theoretical guarantees~\cite{candes2014towards,bhaskar2013atomic}. In particular, we exploit the dual characterization of ANM to construct a certification test: given early-time measurement data and a candidate spectral model, we certify whether that model is compatible with the underlying spectral structure implied by the data. This certification provides a consistency check for reliable forecasting, where correct certification is guaranteed by ANM when the Bohr frequencies are sparse and sufficiently separated, and early-time measurements are low-noise (i.e., the time window where it is possible to obtain reliable measurements from quantum devices and apply error mitigation effectively).
Importantly, the certification framework is independent of the forecasting algorithms, and it can be used to validate candidate models obtained by different forecasting methods, including existing methods in the literature (e.g., \cite{erpenbeck2025compact,kaneko2025forecasting}). 

The main contributions of this work are: 

\begin{itemize} \item We introduce a certification framework based on ANM that tests the consistency of candidate time evolution models used for forecasting quantum observables  (Sec.~\ref{sec:certification_anm}). 
\item We validate the framework on spin-chain time evolutions with 8--20 spin sites, using multiple forecasting algorithms on both noiseless and noisy data (Sec.~\ref{sec:experiments}). 
\end{itemize}

\section{Problem Statement}
\label{sec:model_and_problem}

Consider a quantum system with $n$ qubits (sites) that evolves according to a time-independent Hamiltonian  $H \in \mathbb C^{2^n \times 2^n}$.   Let $| \psi(0) \rangle \in \mathbb C^{2^n}$ be an initial state vector. The quantum time evolution (TE) of the state vector is given by $| \psi(\tau_k) \rangle = e^{-iH \tau_k} |\psi ({0}) \rangle$ where $\tau_k =\Delta (k-1) $ is a discrete time, $\Delta > 0$, and $k \in [m]:=\{ 1, \dots, m \}$ with $m$ the total number of time steps.
The $k$-th measurement of an observable $O \in \mathbb C^{2^n \times 2^n}$ is given by
\begin{align}
y_k := \langle \psi(\tau_k) | O | \psi(\tau_k) \rangle
& = \sum_{f \in \Omega} c(f) e^{i 2 \pi f \tau_k} , \label{eq:trace_o_rho} 
\end{align}
where $c(f) \in \mathbb C$, and $f \in \Omega \subset [-\frac{1}{2},\frac{1}{2}]$ is a Bohr frequency, i.e., normalized difference of the eigenvalues of $H$.

The set $\Omega$ contains only the frequencies with non-zero amplitude, which depend on the spectral information of $H$, the initial state $| \psi(0) \rangle$, and observable $O$. Note that $\Omega$ is generally unknown for large systems since we cannot compute the spectral information of $H$.  
Fig.~\ref{fig:sparse_vs_dense} shows the amplitudes $\mathfrak{R}( c(f))$ for all $f \in \Omega$ for a transverse field Ising model (TFIM) Hamiltonian with two initial states and the same observable. Note that the number of non-zero amplitudes varies, with subfigure (a) having fewer and more spread out frequencies than (b). Also, note that the spectrum is symmetric since observables are real-valued. 

The TE forecasting problem we consider is the following. For $m$ given measurements $y := (y_1,\dots,y_m) \in [-1,1]^m$ and tolerance $\epsilon \ge 0$, find a sequence $x_{k}$ with $k \in [m']$ such that
\begin{align}
\frac{1}{m'} \sum_{k=1}^{m'} | x_k - \langle \psi(\tau_{m+k}) | O | \psi(\tau_{m+k}) \rangle | \le \epsilon. 
\label{eq:objective}
\end{align}
That is, the goal is to predict the $m'$ future values of the time series based on the $m$ available measurements. 

\begin{figure}
\resizebox{\columnwidth}{!}{\input{figures/frequency_separation.tex}}
\caption{Amplitudes $\mathfrak R (c(f))$ for TE under $H = \sum_{i=1}^n \sigma_i^z \sigma_{i+1}^z + \sum_{i=1}^n \sigma_i^x$ with $n=8$ qubits, two initial states, and periodic boundary conditions (site $n+1=1$). The imaginary part of $c(f)$ is zero for both cases. }
\label{fig:sparse_vs_dense}
\end{figure}

\section{Certification via \\Atomic Norm Minimization}

In this section, we briefly review ANM (Sec.~\ref{sec:anm}) and describe how it can be used to certify the frequencies and amplitudes recovered by any spectral estimation algorithm (Sec.~\ref{sec:certification_anm}). 

\subsection{Atomic norm minimization}
\label{sec:anm}

ANM aims to minimize the \emph{atomic norm}: $\| y \|_\mathcal A := \inf \{ t \ge 0 : y \in t \cdot \mathrm{conv}(\mathcal{A})\}$, where $\mathcal A$ is a collection of ``atoms'' (vectors) and $\mathrm{conv}(\mathcal{A})$ its convex hull. The atomic norm can be seen as the continuous analogue of the $\ell_1$ norm, since $\mathcal A$ may contain an uncountable set of atoms corresponding to all possible continuous-frequency sinusoidal components.

For our TE forecasting problem, $\mathcal A$ contains vectors of the form 
\begin{align}
a(f)  := ( e^{i 2 \pi f \tau_1}, \dots, e^{i 2\pi f \tau_m} ),
\label{eq:vector_a}
\end{align}
where $f \in [-\frac{1}{2}, \frac{1}{2}]$, and the ANM finds the sparsest combination of these continuous atoms that match the measurements vector $y$. In some cases, this representation is \emph{unique}, enabling exact recovery of the true frequencies and amplitudes underlying $y$. For the representation to be unique, two conditions must be met: the signal is sparse (i.e., consisting of few frequencies with respect to the number of data points $m$) and the frequencies are sufficiently separated (i.e., $|f_i - f_j| \ge \frac{4}{m}$ for all $f_i, f_j \in \Omega$ with $i \ne j$); see details in \cite[Theorem 1.2]{candes2014towards} and \cite[Theorem 1.1]{tang2013compressed}. 

In practice, we can solve the ANM with atoms of the form in Eq.~\eqref{eq:vector_a} via a semidefinite program (SDP):
\begin{align}
\underset{T, t > 0}{\text{minimize}} \quad \frac{1}{2m}\mathrm{Tr}(T) + \frac{1}{2}t \quad \text{s.t.} \begin{bmatrix} T & y \\ y^* & t \end{bmatrix} \succeq 0,
\label{eq:sdp}
\end{align}
where $T \in \mathbb C^{m \times m}$ is a Hermitian Toeplitz matrix that admits a decomposition $ T = \sum_{f \in F } w(f) a(f) a(f)^*$ (see \cite[Lemma 2.2]{tang2013compressed}), where $w(f) >0$ and $F \subset [-\frac{1}{2},\frac{1}{2}]$ denote the support set of active frequencies. In brief, the constraint in Eq.~\eqref{eq:sdp} ensures that the matrix is positive semi-definite and that $y$ lies in the subspace spanned by vectors $a(f)$. The objective promotes low-rank $T$, thus enforcing sparsity in the frequency domain. Once the optimal $T$ is found, the frequencies can be recovered from $T$ using Prony's method \cite{plonka2014prony}. 

\begin{figure}[t]
\resizebox{\columnwidth}{!}{\input{figures/dual_polynomial.tex}}
\caption{Illustration of the dual polynomial $Q(f)$ (green line) for the setting in Fig.~\ref{fig:sparse_vs_dense} (a). The quantum TE is generated over 10s with a sampling rate of 5 steps per second (i.e., $m = 50$, $\Delta = 1/5)$. The purple bars show $\mathfrak R(c(f))$ as in Fig.~\ref{fig:sparse_vs_dense} (a), where only the $\mathfrak R(c(f))$ larger than 0.005 are shown. The orange markers indicate the sign of $\mathfrak R(c(f))$, which lie on values where $Q(f) = \pm 1$.
}
\label{fig:dual_polynomial}
\end{figure}

Crucially, the correctness of these recovered frequencies can be certified by solving the associated dual problem:
\begin{align}
\underset{q \in \mathbb C^m }{\text{maximize}} \ \langle y, q \rangle  \quad \text{s.t.} \   \sup_{f \in [-\frac{1}{2},\frac{1}{2}]} |Q(f)| \le 1,
\label{eq:dual}
\end{align}
where $Q(f):= \langle a(f),  q \rangle$ is the \emph{dual polynomial}. The solution is certified as \emph{unique} when the dual polynomial reaches its maximum value $|Q(f)| = 1$ precisely for all $f \in \Omega$ and $|Q(f)| < 1 $ otherwise (see \cite[Proposition 2.4]{tang2013compressed}).  An example of this dual certificate is shown in Fig.~\ref{fig:dual_polynomial} for the setting used in Fig.~\ref{fig:sparse_vs_dense} (a), where $Q(f)$ is real-valued and touches $\pm 1$ precisely at the true frequencies  (indicated with markers).

Finally, we note that solving the ANM with the SDP formulation in Eq.~\eqref{eq:sdp} is computationally impractical for a large number of measurements (see example in \cite{supp_material} Supplemental Material, which includes Ref.~\cite{mosek}). In the following section, we show how ANM provides a principled way to certify whether, under sparsity and frequency separation conditions, a candidate model accurately reproduces the true dynamics in Eq. \eqref{eq:trace_o_rho}.

\subsection{Certification pipeline}
\label{sec:certification_anm}

\begin{figure}
\centering
\includegraphics[width=0.9\columnwidth]{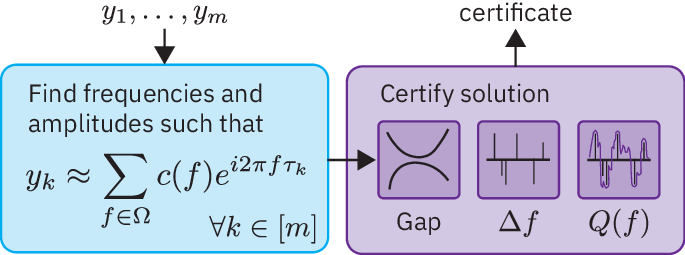}
\caption{Schematic illustration of the proposed certification pipeline. The process verifies frequency and amplitude estimates obtained from any  forecasting algorithm using the dual formulation of ANM.}
\label{fig:cert_pipeline}
\end{figure}

The proposed certification pipeline is illustrated in Fig.~\ref{fig:cert_pipeline}. An algorithm (e.g., ESPRIT \cite{erpenbeck2025compact}) takes as input a vector $y$ with $m$ measurements and returns candidate frequencies $\Omega$ and amplitudes $c(f)$ such that $
y_k \approx \sum_{f \in \Omega} c(f)\, e^{i 2 \pi f \tau_k}, k \in \{1,\dots,m\}$.
These candidates frequencies and amplitudes are then certified by solving the ANM dual problem in Eq.~\eqref{eq:dual}.

\textbf{Certificates.} 
To certify the model, we evaluate three criteria: the duality gap, frequency separation, and the dual polynomial. These three criteria come from ANM (Sec.~\ref{sec:anm}) are sufficient for certifying a model.

The duality gap measures the difference between the primal and dual values (Eqs.~\eqref{eq:sdp} and \eqref{eq:dual}). The gap is always non-negative and, due to strong duality, is zero when both problems are solved optimally. We can obtain an upper bound on the optimal primal objective by constructing $T = \sum_{f \in \Omega} |c(f)| a(f)a(f)^*$, and then setting $t = y^* T^\dagger y$. This construction ensures that the constraints in Eq.~\eqref{eq:sdp} are satisfied, by the Schur complement condition. A lower bound on the dual value can be obtained by solving the dual problem approximately (as discussed later). The duality gap is considered admissible if this is below a prescribed tolerance.

For {frequency separation}, we need to check that the minimum distance between the candidate frequencies in $\Omega$ exceeds $4/m$, which is required for unique recovery in ANM. 

Finally, the {dual polynomial} has to satisfy $|Q(f)| \approx 1$ for all $f \in \Omega$ and $|Q(f)| < 1$, otherwise, within numerical tolerances. This requirement comes from the dual problem in Eq.~\eqref{eq:dual}. Computing the dual polynomial is straightforward once we have a solution to the dual problem. 

\textbf{Solving the dual problem.}
The dual problem in Eq.~\eqref{eq:dual} is infinite-dimensional and must be discretized to make it tractable. If we sample the frequency domain finely, we can transform the problem into a second-order cone program. However, because the measurement vector $y$ is real-valued (which is always the case since observables are real), the dual problem can be further simplified to the linear program
\begin{align}
\max_{q \in \mathbb{R}^m} \ \langle y, q \rangle \quad  \text{s.t.}  \quad -1 \le \langle a(f),  q \rangle \le 1, \ \forall f \in F,
\label{eq:linear_dual_problem}
\end{align}
where $F$ is a finely discretized set of frequencies in $[0, \frac{1}{2}]$, since the frequency spectrum is symmetric. Finally, although discretization introduces a small approximation error, it remains negligible when the frequency grid is sufficiently fine. In the \cite{supp_material} Supplemental Material, we show how selecting different grids affects the accuracy of solving the dual problem.

\section{Experiments}
\label{sec:experiments}

This section evaluates the proposed ANM certification methodology using several forecasting algorithms and spin-chain time evolutions. Importantly, certification is carried out without knowledge of whether the underlying dynamics consist of well-separated Bohr frequencies, since this is something we typically do not know in practice.

\subsection{Setup}
\label{sec:experiment_setup}

\textbf{Dataset.}
We consider quantum TEs generated by 15 spin-chain Hamiltonians (4 TFIM and 11 Heisenberg) with 8–20 qubits and periodic boundary conditions \cite{White1993, Honecker2009, Giamarchi2003, Schulz1986, Affleck1989, Nijs1989, Kurmann1982, Bera2013, Nachtergaele2004, Giampaolo2008} (see details in the \cite{supp_material} Supplemental Material).  For all the experiments, we have used five initial states that are typically  used in the literature \cite{nielsen2010_quantum_computation_quantum_information_10th, brockmann2014_quench_action_approach_releasing_neel, mestyan2015_quenching_xxz_spin_chain_quench,alba2018_entanglement_dynamics_quantum_quenches_generic,zunkovic2018_dynamical_quantum_phase_transitions_spin,richter2020_quantum_quench_dynamics_transversefield_ising} and are easy to prepare in practice (N\'eel, dimerized, paramagnetic, ferromagnetic, and the zero state). Also, we measure six observables ($X$, $Y$, $Z$, $XX$, $YY$, and $ZZ$). This results in a total of 5850 TEs. The TEs are generated using Qutip \cite{johansson2012qutip} for 40 time units with $dt = 0.2$, yielding $m = 200$ measurements per TE. 
A time unit is $t \cdot J$, where $J$ is the coupling strength (interaction energy) in a Hamiltonian.

In the experiments, we consider exact TEs and TEs with added unbiased noise, equivalent to taking 1000 shots. Importantly, the added noise captures uncertainty around the true expectation, and so can be used to model multiple sources of uncertainty, including measurement errors and the residual uncertainty from gate errors after error mitigation. Recall that error mitigation techniques usually trade a reduction in the bias for an increase in variance \cite{cai2023_quantum_error_mitigation}, which is, in essence, the uncertainty around the true value.

\textbf{Spectral estimation algorithms.}
We use five algorithms to compute the frequencies and amplitudes: SDP, Prony, OMP, DMD, and ESPRIT. SDP corresponds to solving the problem in Eq.~\eqref{eq:sdp}. Prony \cite{sauer2018prony} is a method that can provably recover the true frequencies and amplitudes under ideal conditions but is highly sensitive to noise. OMP \cite{chen1994basis, pmlr-v54-locatello17a} is a widely used heuristic in signal processing for sparse signal recovery. DMD \cite{beck2009fast} is the forecasting algorithm used in \cite{kaneko2025forecasting}, and ESPRIT~\cite{paulraj1985_estimation_signal_parameters_rotational_invariance,Roy1989} is a well-known algorithm in spectrum estimation, also applied in observable forecasting \cite{erpenbeck2025compact}. The algorithms are implemented with default or minimal tuning settings, as our focus is on evaluating the certification methodology. A brief description of each algorithm is provided in the \cite{supp_material} Supplemental Material (which includes Refs.~\cite{stoica2005spectral, Wax1985, Golyandina2024, Golyandina2001SSA}), and the code is available online \cite{code}.

\textbf{Forecasting procedure.}
We run every forecasting algorithm (e.g., OMP) with the first $k \in \{10, \dots, 160\}$ measurements and evaluate the forecasting accuracy with the remaining $200-k$ measurements. The forecasting accuracy is measured as indicated in Eq.~\eqref{eq:objective}.

\textbf{Certification.}
For each algorithm run, ANM certification is performed (150 times per algorithm and TE). This is done by discretizing the constraints in Eq.~\eqref{eq:linear_dual_problem} using a frequency grid with 1000 uniformly spaced points in $[0, \frac{1}{2}]$, augmented with the candidate frequencies estimated by the algorithm---to ensure that the dual polynomial is evaluated at the exact/candidate frequencies. Note it is sufficient to consider $[0,\frac{1}{2}]$ instead of $[-\frac{1}{2},\frac{1}{2}]$ since the frequency spectrum is symmetric. To account for numerical inaccuracies, empirically chosen tolerances are applied. Specifically, for exact TEs, the dual polynomial must satisfy $|Q(f)| \ge 0.98$ at all candidate frequencies, and the duality gap must be smaller than 0.05. For TEs with added noise, the thresholds are $|Q(f)| \ge 0.95$ and a duality gap smaller than 0.5. In addition, we only consider amplitudes with magnitude larger than 0.025 for exact TEs and 0.05 for TEs with added noise. The implementation details are available in the \cite{supp_material} Supplemental Material, which includes Ref.~\cite{kingma2014adam}.

\begin{table}[t]
\centering
\footnotesize
\setlength{\tabcolsep}{2pt}
\renewcommand{\arraystretch}{1.2}
\begin{tabular}{c@{\hspace{11pt}}c@{\hspace{2pt}}c@{\hspace{2pt}}c@{\hspace{2pt}}c@{\hspace{2pt}}c@{\hspace{11pt}}c@{\hspace{2pt}}c@{\hspace{2pt}}c@{\hspace{2pt}}c@{\hspace{2pt}}c@{\hspace{2pt}}c@{\hspace{2pt}}c@{\hspace{2pt}}c@{\hspace{2pt}}}
\toprule
 & \multicolumn{5}{c}{Exact TE} & \multicolumn{5}{c}{TE with added noise} \\ 
\cmidrule(lr){2-6} \cmidrule(lr){7-11}
$\epsilon$ & SDP & PNY & OMP & DMD & ESP  &  SDP & PNY & OMP & DMD & ESP  \\ 
\midrule
0.01 & 97.7 & 83.4 & 75.1 & 85.5 & 65.3  & 37.8 &  50.2 & 20.2 & 41.4 & 16.0 \\ 
0.05 & 99.2 & 97.4 & 96.4 & 96.2 & 91.5  & 53.3 & 58.3 & 69.9 & 47.4 & 58.2 \\ 
0.10 & 99.7 & 98.7 & 99.2 & 97.6 & 98.7 & 83.3 & 82.2 & 95.1 & 68.1 & 86.9 \\
\bottomrule
\end{tabular}
\caption{Certification reliability (in percentage) across different forecasting algorithms and tolerances $\epsilon$. The spectral models have been certified with the pipeline in Sec.~\ref{sec:certification_anm}.  }
\label{table:results_SC}
\end{table}

\begin{table}[t]
\centering
\footnotesize
\setlength{\tabcolsep}{2pt}
\renewcommand{\arraystretch}{1.2}
\begin{tabular}{c@{\hspace{11pt}}c@{\hspace{2pt}}c@{\hspace{2pt}}c@{\hspace{2pt}}c@{\hspace{2pt}}c@{\hspace{11pt}}c@{\hspace{2pt}}c@{\hspace{2pt}}c@{\hspace{2pt}}c@{\hspace{2pt}}c@{\hspace{2pt}}c@{\hspace{2pt}}c@{\hspace{2pt}}}
\toprule
 & \multicolumn{5}{c}{Exact TE} & \multicolumn{5}{c}{TE with added noise} \\ 
\cmidrule(lr){2-6} \cmidrule(lr){7-11}
$\epsilon$ & SDP & PNY & OMP & DMD & ESP & SDP & PNY & OMP & DMD & ESP  \\ 
\midrule
0.01  & 16.4 & 15.8 & 11.2 & 16.7 & 18.6 & 5.0 &  6.7 & 7.1 & 7.1 & 7.1 & \\
0.05 & 16.6 & 18.4 & 14.3 & 18.8 & 26.1 &  7.1 & 7.7 & 24.5 & 8.0 & 25.9 & \\
0.10 & 16.7 & 18.6 & 14.7 & 19.1 & 28.2 & 11.1 & 10.9 & 33.4 & 11.5 & 38.8  \\
$\infty$ & 16.8 & 18.9 & 14.9 & 19.6 & 28.5 & 13.3 & 13.2 & 35.1 & 16.9 & 44.6 \\
\bottomrule
\end{tabular}
\caption{Percentage of \emph{successful} certifications over all runs, for different forecasting algorithms and tolerances $\epsilon$. The row with $\epsilon = \infty$ indicates the coverage: the percentage of certifications (successful or not) over all runs.}
\label{table:results_absolute}
\end{table}

\subsection{Results}
\label{sec:experiments_results}
We evaluate forecasting performance using tolerances $\epsilon \in \{0.01, 0.05, 0.1\}$ (see Eq.~\eqref{eq:objective}), for both exact TEs and TEs with added noise. In both cases, we consider the output of a spectral algorithm (i.e., frequencies and amplitudes) \emph{certified} if it meets the criteria in Sec.~\ref{sec:certification_anm}, and we say a certification is \emph{successful} when the forecasting error (as defined in Eq.~\eqref{eq:objective}) is below the desired tolerance $\epsilon$  for the remaining $200-k$ measurements. We define \emph{certification reliability} as the percentage of successful certification across \emph{all certified runs} (successful or not). Namely, how much we can trust that a certified model will generate forecast values with error smaller than or equal to $\epsilon$. In the following, we present certification reliability depending on the number of measurements used to learn the model.

\textbf{Certification reliability (average).}
Table~\ref{table:results_SC} reports the certification reliability for each forecasting algorithm and tolerance $\epsilon$ when algorithms take as input the first $k \in \{10,\dots,160\}$ measurements. For exact TEs, all algorithms achieve high reliability, exceeding 97\% and 91\% for $\epsilon$ equal to 0.1 and 0.05, respectively. With $\epsilon = 0.01$, the performance of all algorithms decreases notably (especially DMD and OMP) with the exception of SDP, which has a reliability of $97.7\%$. The better performance of SDP is because it solves the problem in Eq.~\eqref{eq:sdp} directly. When noise is introduced, reliability decreases across all tolerances and forecasting algorithms, with only OMP achieving a reliability over 95\% for $\epsilon = 0.1$. 

One way to increase the certification reliability is to reduce the coverage (the fraction of certified runs), which can be controlled with the numerical tolerance thresholds used in the certification. However, changing these thresholds creates a trade-off: stricter thresholds improve reliability by excluding borderline cases, but reduce coverage because fewer runs meet the criteria. Table~\ref{table:results_absolute} shows the percentage of successful certifications for different values of $\epsilon$ including the coverage ($\epsilon = \infty$). Observe that the successful certification rates decrease as $\epsilon$ gets smaller and that the values vary across algorithms. For example, Prony and ESPRIT have both a certification reliability of 98.7\% with $\epsilon = 0.1$ and exact TE (see Table \ref{table:results_SC}), but their successful certification rate of ESPRIT is 50\% higher than Prony (Table \ref{table:results_absolute}). In practice, this trade-off implies that certification results should always be reported together with coverage, as high reliability at small $\epsilon$ may reflect only a subset of instances.

\begin{figure}[t]
\resizebox{\columnwidth}{!}{\input{figures/bars.tex}}
\caption{Distribution of the forecasting error upon the first certification (i.e.~using the least amount of measurements necessary for the forecasting model to satisfy certification criteria), with exact and noisy TEs. }
\label{fig:bars}
\end{figure}

\begin{figure}[t]
\resizebox{\columnwidth}{!}{\input{figures/bars_filtered.tex}}
\caption{Distribution of the forecasting error for certification after at least 30 measurements, with exact and noisy TEs.}
\label{fig:bars_filtered}
\end{figure}


\textbf{Certification reliability using a small number of measurements.}
In practice, we are often interested in predicting the TE dynamics using a small number of measurements. 
Fig.~\ref{fig:bars} shows the distribution of the \emph{average} forecasting error (as defined in Eq.~\eqref{eq:objective}) at the \emph{first} certification (the smallest $k \in \{10,\dots,160\}$ that meets the criteria in Sec.~\ref{sec:certification_anm}). Overall, the algorithms exhibit similar behavior. For instance, with exact TEs, Prony and OMP achieve errors below 0.05 and 0.1 in approximately 50\% and 75\% of cases, respectively. For noisy TEs, the proportion of forecasts with errors below 0.05 drops sharply, and both Prony and OMP achieve errors below 0.1 in only about 50\% of cases.

These results contrast with the results in Table~\ref{table:results_SC}, possibly due to insufficient measurements used during spectral estimation or certification. To investigate this, Fig.~\ref{fig:bars_filtered} shows forecasting errors for first certifications issued only after 30 or more measurements. Performance improves noticeably: for exact TEs, Prony achieves errors below 0.05 and 0.1 in 83.4\% and 93.4\% of cases, respectively, while OMP achieves the same thresholds in 76\% and 100\% of cases. For noisy TEs, all algorithms increase the proportion of forecasts with errors exceeding 0.1 (cf.~Fig.~\ref{fig:bars} and Fig.~\ref{fig:bars_filtered}), with OMP achieving errors below 0.1 and 0.15 in 75\% and 97\% of cases, respectively.

These results indicate that a sufficiently large number of measurements is crucial for reliable certification; while the minimum is unknown, monitoring stability, as suggested in \cite[Sec.~III-C]{erpenbeck2025compact}, or setting a minimum measurement threshold (e.g., $k \ge 30$) can significantly enhance performance.

\section{Conclusions}

We have considered the problem of certifying spectral models used for forecasting quantum observables from measurement data. Our proposed certification framework is based on ANM and has guarantees when the time evolution is governed by \emph{a few well-separated Bohr frequencies.} Numerical results show that the certified models can forecast reliably on both noiseless and noisy data, even without prior knowledge of the underlying spectral properties of the time evolution. For instance, OMP with a 0.1 error threshold achieves forecasting reliabilities of 99\% and 95\% on noiseless and noisy data, respectively.

The experiments carried out focus on systems with 8--20 sites (qubits), allowing direct comparison with exact diagonalization, but our certification framework can be applied directly to larger, classically intractable systems. As an example, we show in the \cite{supp_material} Supplemental Material (which includes Refs.~\cite{tiltedIsing1,tiltedIsing2,tiltedIsing3,rudolph2025_pauli_propagation_computational_framework_simulating,SPD}) that for a tilted TFIM with 100 qubits, OMP yields a spectral model that can be certified and used for reliable forecasting. Conversely, certification may fail for systems of arbitrary size if the dynamics are not sufficiently sparse relative to the number of measurements---a limitation that is connected to the spread of quantum information. An interesting direction for future work is to investigate how partial knowledge of the Hamiltonian spectrum can inform the sparsity of the time evolution and, in turn, the number of measurements required for reliable certification.

\section{Acknowledgments} 

We would like to thank the members of the Quantum Information Lab at Tecnun, and  Sergey Bravyi, Ewout van den Berg, Kristan Temme, Niall Robertson, and Nicola Mariella from IBM Quantum. 

This work has been supported by the BasQ strategy of the Department of Science, Universities, and Innovation of the Basque Government through the ``Extrapolation of Von Neumann Dynamics beyond the reach of current Utility Scale Devices (VNDExUSD)'' project.

\bibliography{references}


\clearpage
\onecolumngrid
\appendix
\renewcommand{\thesubsection}{\thesection.\arabic{subsection}}
\begin{center}
    \Large\textbf{Supplemental Material}
\end{center}

\section{Algorithms}
\label{sec:algorithms_appendix}

\subsection{Algorithm for finding frequencies and amplitudes}
\subsubsection{Prony's method}
\label{sec:prony_appendix}

Prony’s method is implemented as described in \cite{sauer2018prony}. 
The method requires that the number of measurements $m$ is larger than $ 2|\Omega|$, 
where $|\Omega|$ denotes the number of frequencies. However, $|\Omega|$ is typically unknown. 
To address this, we execute Prony’s method for a range of candidate values of $|\Omega|$. 
Specifically, given $m$ measurements, we apply Prony’s method using the first $2k$ measurements 
for $k = 1, \dots, \lfloor m/2 \rfloor$. 
For each $k$, we compute the error $\|y - x\|_2$, 
where $x, y \in \mathbb{R}^m$ denote the reconstructed and measured signals, respectively. 
The algorithm selects the model (i.e., the set of frequencies and amplitudes defining $x$) 
that yields the smallest error.

\subsubsection{OMP}

Orthogonal Matching Pursuit (OMP) is implemented as described in \cite[Algorithm~1]{pmlr-v54-locatello17a}, 
using atoms defined by
\begin{align}
a(f,\phi) = (\cos(2\pi f \tau_1 + \phi), \dots, \cos(2\pi f \tau_m + \phi)),
\label{eq:redefined_atom}
\end{align}
where $f \in F := \{0, \tfrac{1}{2L}, \dots, \tfrac{L}{2L}\}$
and $\Phi := \{0, \tfrac{2\pi}{P}, \dots, \tfrac{2\pi(P-1)}{P}\}$, 
with $L = 10000$ and $P = 10$.
We use the atoms in Eq.~\eqref{eq:redefined_atom} instead of those in Eq.~\eqref{eq:vector_a} 
because $y$ is real-valued. Consequently, its frequency spectrum is symmetric, 
and the corresponding amplitudes occur in complex-conjugate pairs.
We make the algorithm terminate after 250 iterations or when the recovered signal, $x$, satisfies $\| y -x\|_2 \le \epsilon$ for $\epsilon = 10^{-2}$. The algorithm returns the frequencies, phases, and amplitude magnitudes of the selected atoms.


\subsubsection{DMD}

The algorithm is implemented according to~\cite{kaneko2025forecasting} and evaluated for successive rank estimates~$r$. Starting from~$r = 1$, the rank is increased incrementally until a satisfactory data fit is achieved or until~$r$ reaches~$m/2$. We consider a data fit is satisfactory if the error is smaller than 0.1. 

\subsubsection{ESPRIT}

ESPRIT (Estimation of Signal Parameters via Rotational Invariance Techniques) was originally proposed for sensor array processing \cite{Roy1989}. The fundamental principle underlying ESPRIT is the rotational invariance of signal subspaces derived from time-shifted observations as outlined in \cite{erpenbeck2025compact}. A systematic treatment can be found in \cite{stoica2005spectral}.

The first step in ESPRIT is to arrange the input signal into a Hankel matrix and then perform its singular value decomposition (SVD). As suggested by the authors of the related Matrix Pencil algorithm \cite{hua1990}, the Hankel matrix width is chosen in the range $n/3$ to $2n/3$ (here $n/2$ by default), where $n$ represents the total number of measurements in the time-series. For further technical details, see \cite{Roy1989, stoica2005spectral, erpenbeck2025compact} and the references therein.

Since we expect a sparse time-series representation, we truncate the SVD spectrum to suppress noisy harmonic components. For sensor array applications, this strategy is well established in the standard literature \cite{Wax1985}. A caveat, however, is that the formulation in \cite{Wax1985} assumes multiple signals impinging on a sensor array, where the matrix columns are modelled as: ``complex (analytic), stationary, and ergodic Gaussian random processes with zero mean and a positive-definite covariance matrix''. In contrast, the columns of a Hankel matrix constructed from a single time series -- as shown in \cite{erpenbeck2025compact}, Eq.~(2) -- are highly correlated (snapshots shifted by one position), and therefore the analysis in \cite{Wax1985} is not directly applicable.

Column decorrelation using subspace projections provides one possible remedy \cite{Golyandina2024, Golyandina2001SSA}. Nevertheless, a simpler empirical approach that works well in practice is to truncate the SVD by removing singular values whose cumulative sum represents only a small, manually set proportion (e.g., 5\%) of the total singular value energy. This is the approach currently adopted here.

\subsection{Algorithm to solve the dual problem in Eq.~\eqref{eq:dual}}

We solve the problem in Eq.~\eqref{eq:dual} with Adam \cite{kingma2014adam}. The objective function is
\begin{align}
f(q) = \langle y, q \rangle + \lambda \sum_{j=1}^{L}\max\{0,|(Aq)_j| - 1\}^2,
\end{align}
where $y \in \mathbb R^m$ is the vector of measurements, and $A \in \mathbb R^{L\times m}$ a matrix with rows equal to the atoms in Eq.~\eqref{eq:vector_a}. The number of atoms corresponds to the discretization of the constraints in the dual problem. In total, $L = 1000 + b$, where $b$ is the number of Bohr frequencies selected by the forecasting algorithm. Specifically, we first select 1000 frequencies uniformly spaced in the interval $[0,\frac{1}{2}]$ and then augment this set with the Bohr frequencies identified by the forecasting algorithm (e.g., OMP).
We set the learning rate equal to 0.1 and run the algorithm for 500 iterations. Parameter $\lambda$ is set to 15.

\section{Hamiltonians and Initial States}
\label{sec:hamiltonians_appendix}

We consider one–dimensional spin-$\tfrac12$ chains. For convenience, we write all Hamiltonians in the \emph{Pauli} convention, i.e., with Pauli matrices $X$, $Y$, and $Z$ acting on the site $k$:
$$
H_{\text{XYZ}} = J_{x} \sum_k X_k X_{k+1} + J_{y} \sum_k Y_k Y_{k+1} + J_{z} \sum_k Z_k Z_{k+1} +
h_x \sum_k X_k + h_y \sum_k Y_k + h_z \sum_k Z_k.
$$
We define the transverse field Ising model (TFIM) as follows:
$$
H_{\text{TFIM}} = J_{z} \sum_k Z_k Z_{k+1} + h_x \sum_k X_k.
$$
Here the index $k$ runs over lattice sites (boundary conditions are specified elsewhere, mostly as ``periodic''), and the couplings $\{J_\alpha\}$ and fields $\{h_\alpha\}$ are given in units where energy scales are $\mathcal{O}(1)$.

Sometimes, spin Hamiltonians are written with spin operators $S_k^\alpha$ rather than Pauli matrices $\sigma_k^\alpha$. Since $S_k^\alpha = \tfrac12\,\sigma_k^\alpha$, our parameters map to the $S$-operator convention as:
$$
J_\alpha^{(S)} = 4\,J_\alpha, \quad h_\alpha^{(S)} = 2\,h_\alpha,
\quad \text{and respectively}, \quad 
J_\alpha = J_\alpha^{(S)}/4, \quad h_\alpha = h_\alpha^{(S)}/2.
$$
Signs are unaffected by this rescaling.

Our study focuses on spin-$\tfrac12$ models in the above Pauli normalization for implementation simplicity. The parameter choices we explore were inspired by Hamiltonians used in diverse physical simulations (see Table~\ref{tab:hamiltonians}). We note, however, that several of those source models were originally formulated for spin-1 or higher; when importing their coefficients into our convention one should apply the normalization map above. Spin-$\tfrac12$ and spin-1 chains can exhibit qualitatively different physics. Because our primary objective here is predictive performance rather than a faithful comparison of phase structures, we do not distinguish between spin-$\tfrac12$ and spin-1 models in the experiments. Results should therefore be interpreted with this caveat in mind.

\begin{table}[hb!]
\centering
\begin{tabular}{ll}
\toprule
\textbf{Initial State\hspace{2em}} & \textbf{Definition} \\
\midrule
Zero & $|0\rangle^{\otimes n}$ \\
N\'eel & $|01\rangle^{\otimes n/2}$\\
Dimerized &  $(|01\rangle - |10\rangle)^{\otimes n/2}$\\
Paramagnetic & $(|0\rangle + |1\rangle)^{\otimes n}$ \\
Ferromagnetic & $|1\rangle^{\otimes n}$ \\
\bottomrule
\end{tabular}
\caption{Initial states and their definitions used in the paper; normalization constants are omitted for brevity.}
\label{tab:init-states}
\end{table}

\begin{table}[ht]
\centering
\begin{tabular}{|p{0.35\textwidth}|p{0.55\textwidth}|}
\hline 
\textbf{Parameters} & \textbf{Comment} \\
\hline
\footnotesize $J_x=0, \,\,\, J_y=0, \,\,\, J_z=-2$,
\hfill\hfill\hfill\linebreak
$h_x=1, \,\,\, h_y=0, \,\,\, h_z=0$ 
& TFIM in the ferromagnetic phase. \\
\hline
\footnotesize $J_x=0, \,\,\, J_y=0, \,\,\, J_z=2$,
\hfill\hfill\hfill\linebreak
$h_x=1, \,\,\, h_y=0, \,\,\, h_z=0$ 
& TFIM in the antiferromagnetic phase. \\
\hline
\footnotesize $J_x=0, \,\,\, J_y=0, \,\,\, J_z=1$,
\hfill\hfill\hfill\linebreak
$h_x=2, \,\,\, h_y=0, \,\,\, h_z=0$ 
& TFIM in the paramagnetic phase. \\
\hline
\footnotesize $J_x=0, \,\,\, J_y=0, \,\,\, J_z=1$,
\hfill\hfill\hfill\linebreak
$h_x=1, \,\,\, h_y=0, \,\,\, h_z=0$ 
& TFIM at the critical point. \\
\hline
\footnotesize $J_x=J_y=J_z=1$,
\hfill\hfill\hfill\linebreak
$h_x=h_y=0, \,\,\, h_z=0.2$ 
& Isotropic Haldane, tiny longitudinal field,
\cite{White1993}. \\
\hline
\footnotesize $J_x=J_y=J_z=1$, 
\hfill\hfill\hfill\linebreak $h_x=h_y=0, \,\,\, h_z=0.4105$ 
& Antiferromagnetic quantum chain at the lower critical field, 
\cite{Honecker2009}. \\
\hline
\footnotesize $J_x=J_y=J_z=1$,
\hfill\hfill\hfill\linebreak $h_x=h_y=0, \,\,\, h_z=0.6$
& Field-induced Tomonaga–Luttinger liquid, \cite{Giamarchi2003}. \\
\hline
\footnotesize $J_x=0.7, \,\,\, J_y=0.7, \,\,\, J_z=1.3$,
\hfill\hfill\hfill\linebreak 
$h_x=0, \,\,\, h_y=0, \,\,\, h_z=0.1$
& Easy-axis XXZ with weak field, \cite{Schulz1986}. \\
\hline
\footnotesize $J_x=1.0, \,\,\, J_y=1.0, \,\,\, J_z=0.6$,
\hfill\hfill\hfill\linebreak 
$h_x=0.2, \,\,\, h_y=0, \,\,\, h_z=0$
& Easy-plane XXZ, transverse probe, \cite{Affleck1989}, \\
\hline
\footnotesize $J_x=J_y=J_z=1$, 
\hfill\hfill\hfill\linebreak 
$h_x=0.3, \,\,\, h_y=0, \,\,\, h_z=0.3$
& Tilted field, symmetry breaking, \cite{Nijs1989}. \\
\hline
\footnotesize $J_x=1.2, \,\,\, J_y=0.8, \,\,\, J_z=1$, 
\hfill\hfill\hfill\linebreak 
$h_x=0.9, \,\,\, h_y=0, \,\,\, h_z=0$
& Anisotropic XYZ + transverse field, \cite{Kurmann1982}. \\
\hline
\footnotesize $J_x=J_y=J_z=1$, 
\hfill\hfill\hfill\linebreak 
$h_x=0, \,\,\, h_y=0, \,\,\, h_z=1$
& Isotropic AFM, intermediate longitudinal field, \cite{Bera2013}. \\
\hline
\footnotesize $J_x=J_y=J_z=-1$, 
\hfill\hfill\hfill\linebreak 
$h_x=0, \,\,\, h_y=0, \,\,\, h_z=0.1$
& Isotropic ferromagnetic Heisenberg chain, tiny field, \cite{Nachtergaele2004}. \\
\hline
\footnotesize $J_x=1, \,\,\, J_y=1, \,\,\, J_z=-0.7$, 
\hfill\hfill\hfill\linebreak 
$h_x=0.15, \,\,\, h_y=0, \,\,\, h_z=0.15$
& Mixed-sign anisotropy, easy-plane exchange but ferromagmetic $J_z$, \cite{Giamarchi2003}. \\
\hline
\footnotesize $J_x=J_y=J_z=1$, 
\hfill\hfill\hfill\linebreak 
$h_x=1.5, \,\,\, h_y=0, \,\,\, h_z=0$
& Isotropic AFM with strong transverse field (near product paramagnet) $J_z$, \cite{Giampaolo2008}. \\
\hline
\end{tabular}
\caption{Spin models and parameters of their Hamiltonians selected for the numerical experiments.}
\label{tab:hamiltonians}
\end{table}

Table~\ref{tab:init-states} summarizes all the initial states used in our numerical simulations. These states are typically used in the literature (to name a few, see~\cite{nielsen2010_quantum_computation_quantum_information_10th, brockmann2014_quench_action_approach_releasing_neel, mestyan2015_quenching_xxz_spin_chain_quench,alba2018_entanglement_dynamics_quantum_quenches_generic,zunkovic2018_dynamical_quantum_phase_transitions_spin,richter2020_quantum_quench_dynamics_transversefield_ising}), making them both representative and practical for our benchmarks.

\clearpage

\section{Additional experiments}

\subsection{Ising model with a tilted field and potential scalability}
\label{sec:app:tilted-tfim}

\begin{figure}[t]
    \begin{tabular}{ccc}
    \includegraphics[width=0.33\textwidth]{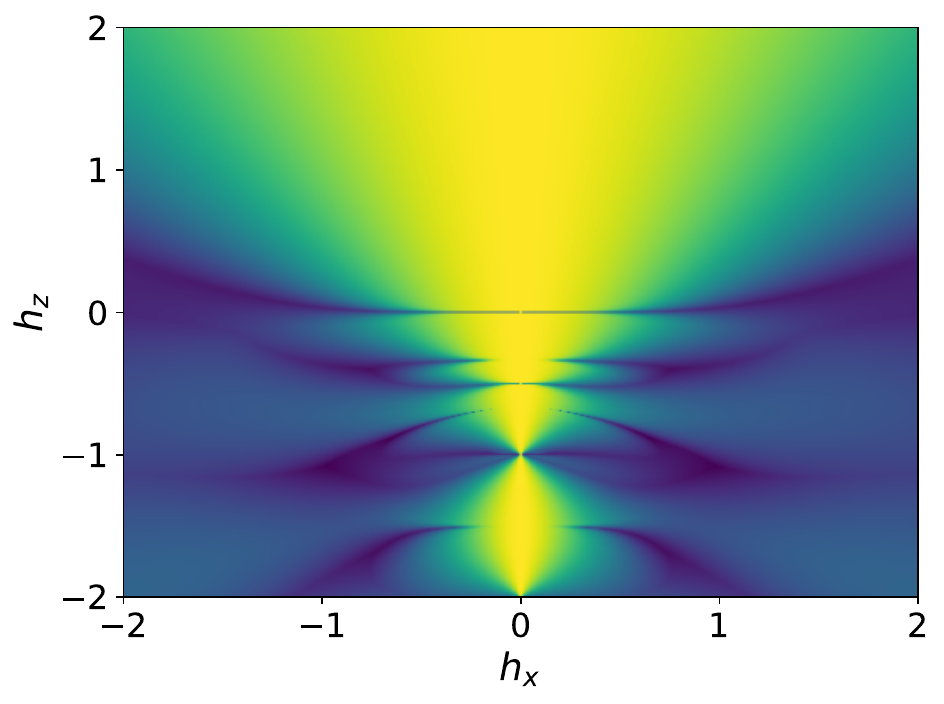}
    & \includegraphics[width=0.33\textwidth]{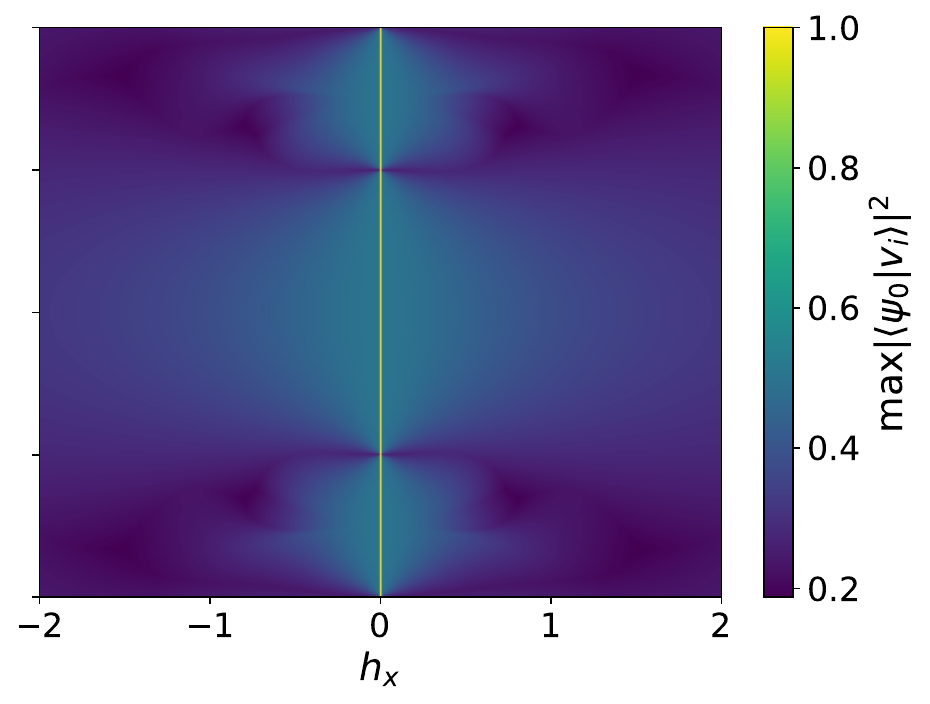} 
    & \includegraphics[width=0.33\textwidth]{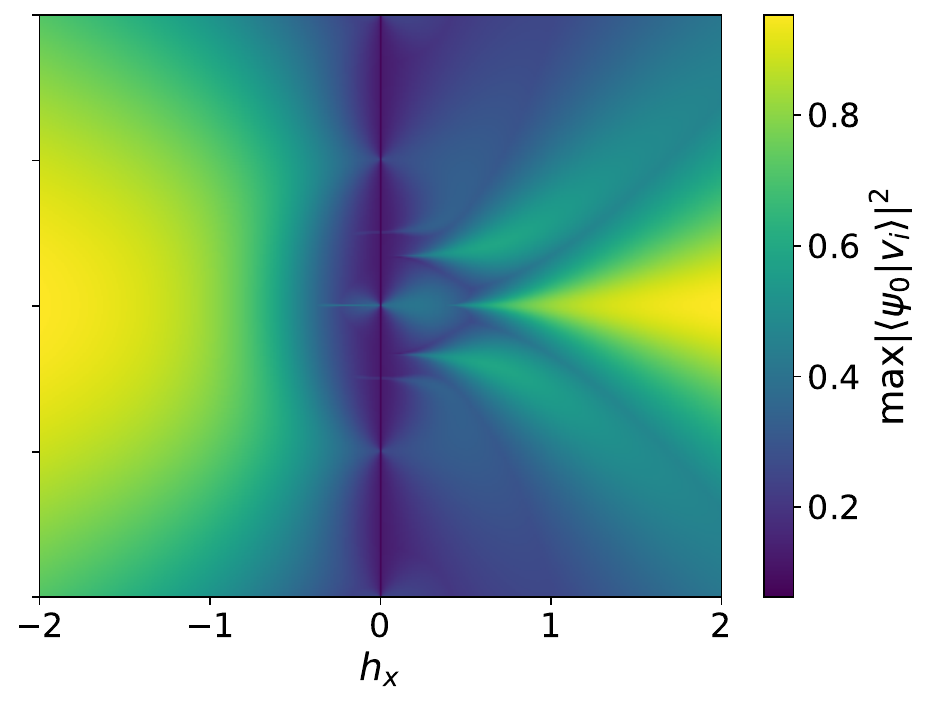} \\
      (a) State $\psi_0 = |0\rangle^{\otimes n}$
    & (b) State $\psi_0 = |01\rangle^{\otimes n/2}$
    & (c) State $\psi_0 = |+\rangle^{\otimes n}$ 
    \end{tabular}
    \caption{Absolute value squared of the $\max$ overlap of an initial state $\psi_0$ with eigenvectors of a Hamiltonian for the Ising model ($J=1.0$) with a tilted field. The overlap is calculated on a $(h_x,h_z)$ plane for $n=4$ sites case. The increase of $n$ leads to shrinking of the overlap, however, a qualitative picture preserves.}
    \label{fig:tilted_tfim_heatmaps}
\end{figure}
In the main text, we extensively tested the proposed certification pipeline for Hamiltonians involving 8--20 spin sites. A natural question is whether this pipeline can scale up to systems with hundreds of sites. Our main discussion is that the success of the certification method depends on the sparsity of the signal and not directly on the system dimensionality. Therefore, here we discuss the potential scalability of our method to systems with a hundred spin sites.

The tilted field Ising model extends the standard Ising model by applying an external magnetic field that is not perfectly aligned with the axes of the model, \textit{i.e.} it includes a combination of a longitudinal and a transversal magnetic fields. The Hamiltonian of such a model can be written as 
\begin{equation}
    H = J\sum_{\langle i,j\rangle}Z_iZ_j + h_x\sum_i X_i + h_z\sum_i Z_i,
\end{equation}
where $h_x,h_z$ refer to the magnitudes of the transverse and longitudinal fields and $J$ refers to the coupling. The importance of the model relies on the introduction of richer physics due to the tilted nature of the applied field, including quantum chaos and integrability breaking. In this regard, the tilted field Ising model is a relevant model to understand how perturbations affect the physics of quantum many-body systems bridging integrable and non-integrable physics, with important consequences in quantum chaos~\cite{tiltedIsing1,tiltedIsing2,tiltedIsing3}.

The interplay between the fields, $h_x$ and $h_z$, allows one to find a sweet spot for a given initial state. Fig~\ref{fig:tilted_tfim_heatmaps} shows the $\max$ overlap of of Hamiltonian's eigenvectors with the initial state $\psi_0$: $\max |\langle \psi_0 | v_i \rangle|^2 \in[0,1]$. When such an overlap equals 1, the initial state is an eigenvector of $H$, and only trivial dynamics can be observed (there is only one frequency contributes to the dynamics). On the other hand, when the overlap $\rightarrow0$, the dynamics becomes far from sparse and the compressed sensing techniques cannot be applied. In this way, we can find a phase diagram like map showcasing Hamiltonian parameters for which the signal will be potentially sparse. Ideal cases to get the sparse dynamics can be found in $(0.8-1.0)$ region of the overlap value. Note that the plots show the $n=4$ case for demonstration purpose; the increase of the number of sites shrinks the areas of large overlap.

Based on Fig.~\ref{fig:tilted_tfim_heatmaps}, we selected a potentially certifiable case when we scale the system to $n=100$ sites. Following this, Fig.~\ref{fig:tilted_tfim_extrapolation} shows a correlator $\langle Z_1(t) Z_2(t) \rangle$ for $n=100$ site Ising model with a tilted field ($J=1.0$, $h_x=0.375, h_z=0.5$) and open boundary conditions, which is potentially sparse Fig.~\ref{fig:tilted_tfim_heatmaps}. Such a combination gives high overlap with the initial state $\psi_0 = |0\rangle^{\otimes n}$, leading to a sparse dynamics that can be learnt and then extrapolated as we showcase in the figure. These results showcsae that our certification/forecasting methods can in fact scale to systems with non-trivial sizes and that the overlap heatmaps of smaller systems can be useful in determining potential extrapolability of bigger systems.

In order to generate the signal in Fig.~\ref{fig:tilted_tfim_extrapolation}, we employed the classical technique usually referred to as Pauli Propagation or Sparse Pauli Dynamics \cite{rudolph2025_pauli_propagation_computational_framework_simulating,SPD} using the PauliPropagation.jl implementation \cite{rudolph2025_pauli_propagation_computational_framework_simulating}. We employed a first order Trotter formula with a $dt=0.2$ to generate $160$ time steps and truncated the Pauli Paths associated to coefficients smaller than $10^{-8}$ and Pauli weights higher than $10$. In order to test convergence, we gradually increased the truncation demands until no improvement was observed in the signal.

\begin{figure}
\resizebox{0.9\textwidth}{!}{\input{figures/100_tilted.tex}}
\caption{Forecasting of the correlator $\langle Z_1(t) Z_2(t) \rangle$ for a $n=100$ sites tilted field Ising model with $J=1.0,h_x=0.375,h_z=0.5$. The red vertical line marks the training interval $[0 ,\dots, 80)$, where we fit the sparse model; the remaining data $[80,\dots,160)$ are reserved for out-of-sample verification. The forecasting algorithm used is OMP.
} 
\label{fig:tilted_tfim_extrapolation}
\end{figure}

\subsection{Two-point correlator example: invalid certification}
\label{sec:two-point-correlator}

\begin{figure}[h]
\centering
    \begin{tabular}{cc}
        \includegraphics[width=0.61\textwidth]{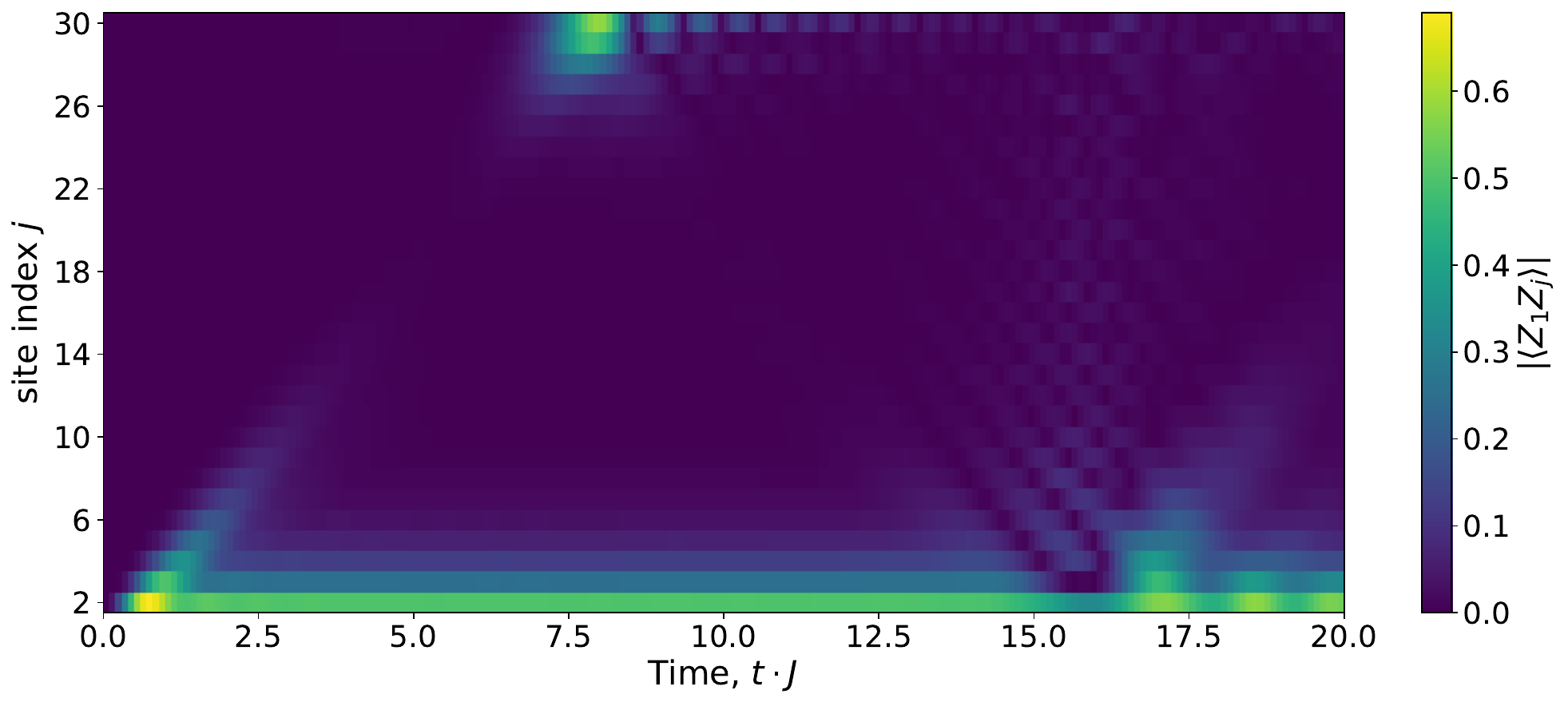} &  \resizebox{0.39\textwidth}{!}{\input{figures/chain_information_speed.tex}} \\
      (a) Lieb--Robinson bound manifestation. & (b) $\langle Z_1(t)Z_{30}(t) \rangle$
    \end{tabular}
    \caption{The two-points correlators $|\langle Z_1(t)Z_j(t) \rangle|$ for $j=2,3,\ldots,n$ for a TFIM with $n=30$ sites (open boundary conditions) and $\psi_0=|+\rangle^{\otimes n}$ initial state.}
    \label{fig:correlator_30qubits_example}
\end{figure}

Quantum information spreads in spin chains with a finite speed, defined by the Lieb--Robinson bound~\cite{anthonychen2023_speed_limits_locality_manybody_quantum}. Fig.~\ref{fig:correlator_30qubits_example}a shows the ``light cone'' heatmap produced by a two-point correlator, $\langle Z_1(t)Z_j(t) \rangle$ with $j=2,3,\ldots,n$, for a TFIM with $n =30$ sites, $\psi_0=|+\rangle^{\otimes n}$ initial state, and open boundary conditions. Correlations first appear for a small value of the site index $j$, then propagate onward in time: $|\langle Z_1(t)Z_j(t) \rangle| \lesssim e^{-(j-vt)}$, where $v$ defines the linear speed of information propagation. 
Fig.~\ref{fig:correlator_30qubits_example}b
 shows that the correlation between the first and the last site in the chain is zero for the first 60 measurements. Hence, measurements of $\langle Z_1(t) Z_n(t) \rangle$ before this time do not contain information about later oscillations (see figure). The figure also shows the forecast obtained with OMP when using only the first 60 measurements (the red vertical line divides the training and forecasting intervals). Observe that the OMP forecast does not capture the oscillations that appear later.

\section{Hyperparameters selection}
\label{sec:hyper}

The hyperparameter selection described in Sec.~\ref{sec:experiment_setup} was carried out empirically using the code in \cite{code}. The code generates an animation that illustrates the time evolution forecast, including how the extrapolation error, the dual polynomial, and the duality gap evolve over time. The selection process was performed on a subset of Hamiltonians, initial states, and observables used in the experiments in Sec.~\ref{sec:experiments}. The algorithms used for selecting the hyperparameters were OMP and Prony since they run faster than the other methods. The choice of hyperparameters can be fine-tuned to each specific time evolution (Hamiltonian, observable, and initial state). However, rather than fine-tuning the parameters for each case, our goal is to find some parameters that work well generally. 

Below, we show an example of the hyperparameter selection for the setting in Fig.~\ref{fig:sparse_vs_dense} (a) with 150 measurements and without noise. Specifically, we show how (i) the forecasting error, (ii) the duality gap, and (iii) the number of dual constraint violations (i.e., $\sum \{ |Q(f)| < \gamma \}$ where $\{ |Q(f)| < \gamma \} = 1$ if $|Q(f) | < \gamma$ and $0$ otherwise, with $\gamma > 0 $ is a threshold and $f$ a frequency in the forecasting model) vary as a function of the number of measurements used. We present three cases corresponding to different thresholds for filtering amplitude magnitudes (0.05, 0.025, and 0.0125) and thresholds for the dual polynomial values (0.9, 0.95, and 0.99). Additionally, for each experiment we use different number of frequencies when solving the dual problem (100, 1000, 10,000 and 100,000).

Fig.~\ref{fig:hyper_0} (left) shows the forecasting error and duality gap for different frequency grids when solving the dual problem (100–100,000) and when the amplitude magnitudes of the model are filtered to be larger than 0.05. Observe from the figure that the duality gap changes depending on the number of frequencies used. With 100 frequencies, the duality gap is large and does not decrease as the number of measurements increases. With 1000 frequencies, we observe that the duality gap decreases over time along with the forecasting error, which is the desired behaviour. With 10,000 and 100,000 frequencies, we observe a similar behaviour as with 1000 frequencies; however, note that the duality gap is slightly negative at some points. This is due to numerical issues arising from the discretization of the dual constraints.

Fig.~\ref{fig:hyper_0} (right) shows the number of frequencies in the model that do not satisfy $|Q(f)| \ge 0.9$ (i.e., $\sum \{|Q(f)| < 0.9\}$). Recall a forecasting model is \emph{not} certified if  $\sum \{|Q(f)| < 0.9\} \ne 0$. Observe from the figure that there is a strong correlation between the number of frequencies that violate the constraint and the forecasting error and duality gap. Specifically, with 100 frequencies, we find that the model is never certified since $\sum \{|Q(f)| < 0.9\} > 0$, whereas for the other cases $\sum \{|Q(f)| < 0.9\} = 0$ occurs around measurement 60.

Fig.~\ref{fig:hyper_1} and \ref{fig:hyper_2} show behaviour similar to Fig.~\ref{fig:hyper_0} in terms of forecasting error and duality gap, but with amplitude magnitude thresholds 0.025 and 0.0125 respectively. The behavior of constraint violations changes. In Fig.~\ref{fig:hyper_1}, the number of measurements required to certify the model increases from around 60 to almost 80 due to the stricter thresholds (0.95), while in Fig.~\ref{fig:hyper_2} the model is never certified due to the threshold being too strict (0.99). The latter is a combination of allowing the model to use frequencies with smaller amplitude magnitudes, while also imposing a stricter threshold on $|Q(f)|$.

Finally, note that we can control the \emph{coverage} through numerical thresholds, where coverage is defined as the fraction of certified time evolutions over all runs. The number of frequencies used affects the accuracy with which we can solve the dual problem and, consequently, the duality gap criterion we use in the certification. Similarly, the thresholds applied to amplitude magnitudes and dual-constraint violations determine whether or when $\sum \{|Q(f)| < \gamma\} = 0$ for some $\gamma > 0$, which is also necessary for certification. 

\begin{figure}[h!]
\begin{tabular}{cc}
\resizebox{0.45\textwidth}{!}{\input{figures/hyper_0.tex}}
& 
\resizebox{0.45\textwidth}{!}{\input{figures/hyper_0_poles.tex}}
\end{tabular}
\caption{OMP with frequencies that have amplitude magnitude than 0.05 and $|Q(f)| \ge 0.9$ for all selected frequencies. The horizontal gray line indicates the zero value. }
\label{fig:hyper_0}
\end{figure}


\begin{figure}
\begin{tabular}{cc}
\resizebox{0.45\textwidth}{!}{\input{figures/hyper_1.tex}}
& 
\resizebox{0.45\textwidth}{!}{\input{figures/hyper_1_poles.tex}}
\end{tabular}
\caption{OMP with frequencies that have amplitude magnitude than 0.025 and $|Q(f)| \ge 0.95$ for all selected frequencies. The horizontal gray line indicates the zero value.}
\label{fig:hyper_1}
\end{figure}


\begin{figure}
\begin{tabular}{cc}
\resizebox{0.45\textwidth}{!}{\input{figures/hyper_2.tex}}
& 
\resizebox{0.45\textwidth}{!}{\input{figures/hyper_2_poles.tex}}
\end{tabular}
\caption{OMP with frequencies that have amplitude magnitude than 0.0125 and $|Q(f)| \ge 0.99$ for all selected frequencies. The horizontal gray line indicates the zero value.}
\label{fig:hyper_2}
\end{figure}

\clearpage
\section{SDP Running Times with MOSEK}
\label{sec:sdp_running_times}

Fig.~\ref{fig:sdp_times} shows the time it takes MOSEK \cite{mosek} (a commercial SDP solver) to solve the problem in Eq.~\eqref{eq:sdp}. The SDP is run on an Intel i7-14700 CPU with 32 GB of memory. 

\begin{figure}[h]
\resizebox{0.45\textwidth}{!}{\input{figures/sdp_times.tex}}
\caption{Time to solve the SDP in Eq.~\eqref{eq:sdp} with MOSEK \cite{mosek} for different number of measurements. }
\label{fig:sdp_times}
\end{figure}

\end{document}